\documentclass[a4paper,11pt]{article}

\usepackage[width=15cm,height=23cm]{geometry}
\usepackage{amsfonts,amssymb}

\usepackage[OT2,OT1]{fontenc}
\usepackage[ansinew]{inputenc}

\def\cyr{\fontencoding{OT2}\fontfamily{wncyr}\selectfont}

\def\G{\Gamma}

\begin{document}
\par\noindent
\rightline{FR-PHENO-2010-012}
\rightline{February 2010}

\vskip 4 truecm
\bf
\Large
\bf
\centerline{
Renormalization of the Yang-Mills theory}
\centerline{ in the
ambiguity-free gauge.}

\normalsize
\rm

\large
\rm
\begin{center}
\vskip 0.7 truecm
A.~Quadri$^{a,b}$\footnote{e-mail: {\tt andrea.quadri@mi.infn.it}},
A.~A.~Slavnov$^{c}$\footnote{e-mail: {\tt slavnov@mi.ras.ru}}
\end{center}

\normalsize
\begin{center}
$^a$
Physikalisches Institut\\
Albert-Ludwigs Universit\"at Freiburg\\
Hermann-Herder-Strasse 3a,\\
D-79104 Freiburg i.Br., Germany\\
$^b$
Dip. di Fisica, Universit\`a degli Studi di Milano\\
via Celoria 16, I-20133 Milano, Italy\\
$^c$
Steklov Mathematical Institute\\
Gubkina st.8, Moscow, Russia
\end{center}

\vskip 0.8 truecm

\begin{abstract}
The renormalization procedure for the
Yang-Mills theory in the gauge free of the Gribov ambiguity is
constructed. It is shown that all the ultraviolet infinities may be
removed by renormalization of the parameters entering the classical
Lagrangian and the local redefinition of the fields.
\end{abstract}

\vskip 5 truecm

\newpage

\section{Introduction} \label{sec.1}

A problem of unambiguous quantization of nonabelian gauge theories
remains unsolved. Although in the framework of perturbation theory
a consistent quantization procedure was formulated by L.Faddeev
and V.Popov~\cite{FP}
and B.DeWitt~\cite{DW}, for large fields as was pointed out by V.Gribov~\cite{VG} the Coulomb
gauge condition $\partial_iA_i=0$ normally used in the process of
quantization does not choose a unique representative in the class
of gauge equivalent configurations. This result was later
generalized by I.Singer~\cite{IS} to arbitrary covariant gauge.
At present it is not clear if this problem
leads to serious physical consequences and different proposals
how to overcome this difficulty were formulated
(see for example \cite{DZ}). However in our opinion
a satisfactory solution has not been found.

The Gribov ambiguity  usually  arises when a gauge condition
includes a differential operator, which leads to the existence
of nontrivial solutions of the equation $L(A_{\mu}^{\Omega},\varphi^{\Omega})=0$,
considered as the equation for the gauge function $\Omega$
at the surface $L(A, \varphi)=0$. So
the most direct way to avoid this difficulty would be
to consider so called algebraic gauges
$\tilde{L}(A_{\mu}, \varphi)=0$, where the operator
$\tilde{L}$ does not involve a differentiation.
Example of such gauges is given by the condition $nA=0$, where $n$ is some
constant vector. Such gauges are known to be ghost free,
but they violate explicitly the Lorentz
invariance, resulting in the serious complications in analysis of the model.
Moreover they lead to some additional problems which will not be discussed here.

The BRST quantization~\cite{KO} avoids the problem of the
gauge-fixing ambiguity. However in
this case the gauge invariance is broken even at the classical level
and one can prove the independence of observables on the gauge
chosen only in the framework of perturbation theory.

On the other hand studies of QCD require a reliable
gauge invariant method, valid beyond
perturbation theory. It is highly desirable
that this method preserves the manifest Lorentz
invariance.

According to the common lore the quantization of the gauge invariant
Yang-Mills theory always leads to the above mentioned problems.
Algebraic gauges violate Lorentz invariance, whereas differential
gauges are plagued by the Gribov ambiguity.

Recently a modified formulation of the Yang-Mills theory
was proposed, which admits Lorentz
invariant algebraic gauge conditions \cite{Sl1}.
Contrary to the standard formulation in these
gauges the ghost field Lagrangian is manifestly
gauge invariant. A consistent quantization
procedure
was developed on the basis of this formulation \cite{Sl2}.
However these gauges suffered from a new problem.
Although the degree of divergency of arbitrary
diagrams was limited, the number of primitively
divergent diagrams was infinite and the standard
perturbative renormalization of the model failed.
Formally one could pass from the  gauge  proposed in \cite{Sl2} to the standard differential
gauges like $\partial_{\mu}A_{\mu}=0$, in which the
theory is manifestly renormalizable, but this transition is legitimate
only in the framework of perturbation theory.
Of course for perturbative renormalization it is sufficient, but if one is
planning to consider big fields as well, it is very useful
to have a formulation which makes
sense both in the framework of perturbation theory and beyond it.
Moreover many modern attempts to study nonperturbative
behaviour of QCD use some resummation of perturbation series.
It requires a formulation of the model which may be used
beyond perturbation theory and is
perturbatively renormalizable.
Such a formulation, which allows to perform a consistent quantization
irrespectively of the validity of the perturbation theory was proposed in
the papers \cite{Sl1}, \cite{Sl2}.
 However renormalizability of the
theory in this approach was not obvious. In the present paper we show that
the proposed theory is indeed renormalizable and therefore may serve as a
starting point for non-perturbative approximations. In the framework of
perturbation theory it produces the results coinciding with the standard
formalism. Examples of the perturbative calculations confirming this
statement were presented earlier in the paper \cite{Bar}.

Having this in mind in the present paper we studied the
renormalization procedure for the ambiguity free formulation of the
Yang-Mills theory.     We show that
all the ultraviolet divergencies may be removed by introducing a
finite number of counterterms, supplemented by suitable
fields redefinitions which preserve gauge invariance.

The paper is organized as follows. In the Section \ref{sec.2} we
introduce the model and discuss its symmetries and the problem of
unitarity. In the Section \ref{sec.3}    we discuss the problem of renormalization of the Yang-Mills theory in the ambiguity free gauge.   In the
Section \ref{sec.4} the renormalization of the model is described
and the renormalized Lagrangian is presented. In the Section
\ref{sec.5} the problem of gauge independence of the renormalized
theory is considered and equivalence of the present model to the
standard Yang-Mills theory is proven. In the appendix \ref{app.B}
the symmetry preserving renormalization procedure which includes the
field redefinition is explicitely constructed. Finally possible
applications are briefly discussed.

\section{The effective action and its symmetries}\label{sec.2}

To save the place we consider the model invariant with respect to $SU(2)$ gauge group.
Generalization to $SU(N)$ groups makes no problems.

To illustrate the main idea we consider firstly
the classical Lagrangian of Yang-Mills theory supplemented by
gauge-invariant couplings of the scalar fields $\varphi, \chi, b, e$
\cite{Sl1,Sl2}
\begin{eqnarray}
{\cal L} = -\frac{1}{4} F^a_{\mu\nu} F^a_{\mu\nu}
+ (D_\mu \varphi)^* D_\mu \varphi
-  (D_\mu \chi)^* D_\mu \chi  -
[(D_{\mu}b)^*(D_{\mu}e)+(D_{\mu}e)^*(D_{\mu}b)] \label{0}
\end{eqnarray}
Here $F_{\mu\nu}^a$ is the standard curvature tensor for the Yang-Mills field. The scalar fields
$\varphi, \chi, b, e$ form the complex $SU(2)$ doublets parametrized
as follows:
\begin{equation}
 \Phi=\left( \frac{i\Phi_1+\Phi_2}{\sqrt{2}}, \frac{\Phi_0-i\Phi_3}{\sqrt{2}}\right)
\label{2}
\end{equation}
where $\Phi$ denotes any of doublets. The fields $\varphi$ and $\chi$ are commuting,  and their components $\varphi^{\alpha}, \chi^{\alpha}$ are hermitean. The
fields $e$ and $b$ are anticommuting.  By definition the components $e^{\alpha}$ are hermitean, and the components $b^{\alpha}$ are antihermitean.  In the eq.(\ref{0}) $D_{\mu}$ denotes the usual covariant
derivative.

As explained in \cite{Sl1}, if the asymptotic states do not
contain the exitations corresponding to the scalar fields
$\varphi,\chi, b, e$ one may perform in the path-integral
\begin{eqnarray}
S = \int \exp \Big ( i \int \, {\cal L} ~ dx \Big ) \delta(\partial_i A_i)
d\mu
\label{1.bis}
\end{eqnarray}
the integration over the scalar fields. Then one gets the
factor $(|D^2|)^{-2}$ from the integration over the commuting fields
$\varphi$ and $\chi$ and the factor $(|D^2|)^{2}$ from the
integration over the anticommuting fields $b,e$, so finally one ends
up with the path-integral of the usual Yang-Mills theory in the
Coulomb gauge.

Now we consider a different Lagrangian, which may be obtained
from (\ref{0}) by
 the following shift
 of the commuting scalar fields
\begin{equation}
\varphi\rightarrow\varphi-g^{-1} \hat{m};  \quad \chi\rightarrow \chi+g^{-1} \hat{m} \, .\label{4}
\end{equation}
The constant field $\hat{m}$ has a form
\begin{equation}
\hat{m}=(0,m) \, . \label{3}
\end{equation}
Then we obtain
the classical Lagrangian
\begin{eqnarray}
L=- \frac{1}{4}F_{\mu\nu}^aF_{\mu\nu}^a+(D_{\mu}\varphi)^*(D_{\mu}\varphi)-(D_{\mu}\chi)^*
(D_{\mu}\chi)\nonumber\\
-g^{-1}[(D_{\mu}\varphi)^*+(D_{\mu}\chi)^*](D_{\mu}\hat{m})-g^{-1}(D_{\mu}\hat{m})^*[D_{\mu}\varphi+D_{\mu}\chi]\nonumber\\
-[(D_{\mu}b)^*(D_{\mu}e)+(D_{\mu}e)^*(D_{\mu}b)] \label{1}
\end{eqnarray}
 Note that due to the negative sign of the $\chi$ field
kinetic term, this field
possesses negative energy. This is crucial in order to ensure
the cancellation of the quadratic terms in $m$ in
eq.(\ref{1}) and therefore to provide a zero mass for the
Yang-Mills fields. The factor $g^{-1}$ in the shift in eq.(\ref{4})
is chosen in such a way that the scalar-gauge fields bilinears do
not depend on $g$, and $g$ only enters into interaction vertices in
eq.(\ref{1}).

One may think that the massive parameter $m$ enters into expressions
for observable gauge-invariant expectation values. We shall prove
however that the observables do not depend on this parameter. A
similar situation occurs in the Higgs model, if one works in the
unitary gauge. To renormalize the theory at any given order of
perturbation expansion one has to introduce a number of parameters
(massless and massive). However the observable gauge-invariant
quantities do not depend on these parameters, provided the
renormalization preserves gauge invariance. The easiest way to see
that is to pass to some manifestly renormalizable gauge (for example
the Lorentz gauge $\partial_{\mu}A_{\mu}=0$).

It is worth to notice that although the Lagrangian (\ref{1})
may be obtained from the gauge invariant Lagrangian describing the
interaction of the Yang-Mills field with the scalars $(\varphi^\pm,
b, e)$ by the shift (\ref{4}), the theory described by this
Lagrangian may be absolutely different w.r.t. the unshifted one.
Shift of the fields by a constant in general results in a new
theory, inequivalent to the original one. A well-known example is
given by the Higgs model. To prove the equivalence of our theory to
the ordinary Yang-Mills theory one has to prove the decoupling of
all the fields $(\varphi^\pm, b, e)$, temporal and longitudinal
Yang-Mills quanta from the three dimensionally transversal
Yang-Mills quanta. For unrenormalized theory it has been done in the
papers \cite{Sl1,Sl2}. In this paper we shall prove it for the
renormalized theory in the ambiguity free gauge.


The Lagrangian (\ref{1}) is invariant with respect to the following gauge transformations
inherited from the gauge symmetry  of
${\cal L}$ in eq.(\ref{0})
\begin {eqnarray}
\delta A^a_{\mu}= \partial_{\mu}\eta^a+g\epsilon^{abc}A^b_{\mu}\eta^c\nonumber\\
\delta \varphi^0_+= -\frac{g}{2} \varphi_+^a \eta^a\nonumber\\
\delta \varphi^0_-=-\frac{g}{2} \varphi_-^a \eta^a\nonumber\\
\delta \varphi_+^a=\frac{g}{2} \epsilon^{abc}\varphi_+^b\eta^c+
 \frac{g}{2} \varphi_+^0 \eta^a\nonumber\\
\delta \varphi^-_a=m\eta^a+\frac{g}{2} \epsilon^{abc}\varphi_-^b \eta^c+\frac{g}{2} \varphi_-^0\eta^a\nonumber\\
\delta b^a=\frac{g}{2}  \epsilon^{abc}b^b \eta^c+\frac{g}{2}b^0\eta^a\nonumber\\
\delta e^a=\frac{g}{2} \epsilon^{abc}e^b\eta^c+\frac{g}{2}e^0 \eta^a\nonumber\\
\delta b^0=- \frac{g}{2}b^a \eta^a\nonumber\\
\delta e^0= -\frac{g}{2}e^a \eta^a \label{5}
\end{eqnarray}
Here the obvious notations
$$
\varphi^{\alpha}_{ \pm}= \frac{\varphi^{\alpha} \pm \chi^{\alpha}}{\sqrt{2}}
$$
are introduced.

This Lagrangian is also invariant with respect to the supersymmetry transformations
\begin{eqnarray}
\delta \varphi^a_-=- b^a\nonumber\\
\delta \varphi^0_-=- b^0\nonumber\\
\delta e^a=\varphi^a_+\nonumber\\
\delta e^0=\varphi^0_+\nonumber\\
\delta b=0 \nonumber \\
\delta \varphi_+ = 0
\label{6}
\end{eqnarray}

This invariance plays a crucial role in the proof of the equivalence of the model described by
the Lagrangian (\ref{1}) to the standard Yang-Mills theory. It was shown in the papers
\cite{Sl1, Sl2} that it provides the unitarity of the scattering matrix in the subspace
which includes only three dimensionally transversal components of the Yang-Mils field.


The field $\varphi^a_-$ is shifted under the gauge transformation by an arbitrary function $m
\eta^a$. It allows to impose Lorentz invariant algebraic gauge condition $\varphi^a_-=0$.

However imposing the Lorentz invariant gauge condition $ \varphi^a_-=0$ does not solve the problem of ambiguity completely. As it follows from the eq.(\ref{5}) the field $\varphi^a_-$ satisfying the condition $\varphi^a_-=0$ is transformed by the gauge transformation to $\varphi'^a_-=(m+ \frac{g}{2} \varphi^0_-)\eta^a$. For some $x$ the factor $(m+ \frac{g}{2} \varphi^0_-(x))$ may vanish, leading to nonuniqueness of the gauge fixing.

To avoid the problem of ambiguity completely we redefine the fields entering the Lagrangian (\ref{1}) as follows
\begin{eqnarray}
\varphi^0_-=\frac{2m}{g}(\exp\{\frac{gh}{2m}\}-1); \quad
\varphi^a_-=\tilde{M}\tilde{\varphi}^a_-\nonumber\\
\varphi^a_+=\tilde{M}^{-1}\tilde{\varphi}^a_+; \quad \varphi^0_+=
\tilde{M}^{-1}\tilde{\varphi}^0_+\nonumber\\
e= \tilde{M}^{-1} \tilde{e}; \quad b= \tilde{M} \tilde{b}
\label{6a}
\end{eqnarray}
 where
\begin{equation}
\tilde{M}=1+ \frac{g}{2m}\varphi^0_-  = \exp\{\frac{gh}{2m}\} \label{6b}
\end{equation}
The new Lagrangian has the form
\begin{eqnarray}
\tilde{L}=-\frac{1}{4}F^a_{\mu \nu}F^a_{\mu \nu}+ \partial_{\mu}h \partial_{\mu} \tilde{\varphi}^0_+- \frac{g}{2m}\partial_{\mu}h \partial_{\mu}h \tilde{\varphi}^0_+\nonumber\\+m \tilde{\varphi}^a_+ \partial_{\mu}A_{\mu}^a-[((D_{\mu} \tilde{b})^*+ \frac{g}{2m} \tilde{b}^*\partial_{\mu}h)(D_{\mu} \tilde{e}- \frac{g}{2m} \tilde{e}\partial_{\mu}h)+h.c.]\nonumber\\
+ \frac{mg}{2}A_{\mu}^2 \tilde{\varphi}^0_++ g \partial_{\mu}h A_{\mu}^a \tilde{\varphi}^a_+ \ldots
\label{6c}
\end{eqnarray}
Here $ \ldots$ denote the terms $\sim \tilde{\varphi}^a_-$, which are obviously polynomial.

The Lagrangian (\ref{6c}) by construction is invariant with respect to the gauge transformations generated by the transformations (\ref{5}) after the change (\ref{6a}):
\begin{eqnarray}
\delta A^a_{\mu}= \partial_{\mu}\eta^a+g\epsilon^{abc}A^b_{\mu}\eta^c\nonumber\\
\delta \tilde{\varphi}^0_+= -\frac{g}{2} \tilde{\varphi}_+^a \eta^a- \frac{g^2}{4m}\tilde{\varphi}^0_+ \tilde{\varphi}^a_-\eta^a\nonumber\\
\delta h=- \frac{g}{2} \eta^a\tilde{\varphi}^a_-\nonumber\\
\delta \tilde{\varphi}_+^a=\frac{g}{2} \epsilon^{abc}\tilde{\varphi}_+^b\eta^c+
 \frac{g}{2} \tilde{\varphi}_+^0 \eta^a- \frac{g^2}{4m}\tilde{\varphi}_+^a \tilde{\varphi}^b_- \eta^b \nonumber\\
\delta \tilde{\varphi}_-^a=m\eta^a+\frac{g}{2} \epsilon^{abc}\tilde{\varphi}_-^b \eta^c+\frac{g^2}{4m}\tilde{\varphi}_-^a \tilde{\varphi}^b_- \eta^b \nonumber\\\tilde
\delta \tilde{b}^a=\frac{g}{2} \epsilon^{abc}\tilde{b}^b \eta^c+\frac{g}{2} \tilde{b}^0\eta^a+\frac{g^2}{4m} \tilde{\varphi}_-^b
\eta^b \tilde{b}^a\nonumber\\
\delta \tilde{e}^a=\frac{g}{2} \epsilon^{abc}\tilde{e}^b\eta^c+\frac{g}{2}\tilde{e}^0 \eta^a- \frac{g^2}{4m} \tilde{\varphi}_-^b
\eta^b \tilde{e}^a\nonumber\\
\delta \tilde{b}^0=- \frac{g}{2}\tilde{b}^a \eta^a+ \frac{g^2}{4m}\tilde{\varphi}_-^b
\eta^b \tilde{b}^0\nonumber\\
\delta \tilde{e}^0= -\frac{g}{2} \tilde{e}^a \eta^a- \frac{g^2}{4m}\tilde{\varphi}_-^b \eta^b \tilde{e}^0 \label{6d}
\end{eqnarray}

At the surface $\tilde{\varphi}^a_-=0$ the gauge variation of the field $\tilde{\varphi}^a_-$ is equal to $m \eta^a$ and therefore the condition $\tilde{\varphi}^a_-=0$ chooses the unique representative in the class of the gauge equivalent configurations.

Obviously the Lagrangian (\ref{6c}) is also invariant with respect to the supersymmetry transformations generated by the transformations (\ref{6}) after the change (\ref{6a}):
\begin{eqnarray}
\delta h=- \tilde{b}^0 \nonumber\\
\delta \tilde{\varphi}_a^-=- \tilde{b}^a+ \frac{g}{2m}\tilde{\varphi}^a_-\tilde{b}^0\nonumber\\
\delta \tilde{\varphi}_a^+=- \frac{g}{2m} \tilde{b}^0 \tilde{\varphi}^a_+\nonumber\\
\delta \tilde{\varphi}^0_+=-\frac{g}{2m} \tilde{b}^0 \tilde{\varphi}^0_+\nonumber\\
\delta \tilde{e}^a=\tilde{\varphi}^a_++ \frac{g}{2m} \tilde{e}^a \tilde{b}^0\nonumber\\
\delta \tilde{e}^0=\tilde{\varphi}^0_++\frac{g}{2m} \tilde{e}^0 \tilde{b}^0\nonumber\\
\delta \tilde{b}^a= -\frac{g}{2m} \tilde{b}^a \tilde{b}^0\nonumber\\
\delta \tilde{b}^0=0
\label{6e}
\end{eqnarray}

Note  however that imposing the gauge condition
 $\tilde{\varphi}^a_-=0$  we break the invariance of the effective
action with respect to the supersymmetry transformation (\ref{6e}). To overcome this difficulty
we consider in more details the effective action.

A canonical quantization in the gauge    $\tilde{\varphi}^a_-=0$  requires introduction of ultralocal
ghosts. So the gauge fixing is introduced by adding to the action the term
\begin{equation}
s\int d^4x \bar{c}^a\tilde{\varphi}^a_-= \int d^4x (\lambda^a\tilde{\varphi}^a_- - \bar{c}^aM^{ab}c^b) \label{7a}
\end{equation}
where
\begin{equation}
M^{ab}= \delta^{ab}m+ \frac{g}{2}\varepsilon^{acb}
\tilde \varphi^c_-+ \frac{g^2}{4m}
\tilde{\varphi}^a_- \tilde{\varphi}^b_- \label{8a}
\end{equation}

Here $s$ is the BRST operator.
In order to derive its action from eq.(\ref{6d}) some
care is needed with the fermionic fields.
One replaces first $\eta^a
\rightarrow \epsilon c^a$, where $\epsilon$ is a constant anticommuting parameter, and then drops
$\epsilon$ once it has been moved to the left.
Moreover one sets
%
\begin{eqnarray}
(sc)^a=-  \frac{g}{2}\varepsilon^{abc}c^bc^c\nonumber\\
(s \bar{c})^a=  \lambda^a\nonumber\\
(s\lambda)^a=0 \, .\label{9}
\end{eqnarray}
 The gauge fixed action is obviously invariant with respect to this BRST
transformation. It leads to some relations satisfied by the one particle irreducible diagrams,
which will be discussed later.

However, as it was mentioned above, due to the presence of the term $\lambda^a \tilde{\varphi}^a_-$ this
action is not invariant with respect to the supersymmetry transformation (\ref{6e}), which
provides the physical unitarity of the model. As the transition from one gauge to the other one
may be achieved by a gauge transformation, and in the gauge  $\partial_{\mu}A_{\mu}=0$ the
effective action is invariant with respect to the  supertransformation (\ref{6e}), in the gauge
   $\tilde{\varphi}^a_-=0$ it also must be invariant with respect to some supertransformation.

To obtain this transformation we note that the Lagrangian
eq.(\ref{6c})  which is invariant with
respect to the BRST transformation corresponding to the gauge transformations eq. (\ref{6d})  and
with respect to the supersymmetry transformations eq.(\ref{6e}) is also invariant with respect to
the simultaneous change of the fields combining these two transformations:
\begin{eqnarray}
\delta A^a_{\mu}= \partial_{\mu}c^a+g\epsilon^{abc}A^b_{\mu}c^c\nonumber\\
\delta \tilde{\varphi}^0_+= -\frac{g}{2} \tilde{\varphi}_+^a c^a- \frac{g^2}{4m}\tilde{\varphi}^0_+ \tilde{\varphi}^a_-c^a -\frac{g}{2m} \tilde{\varphi}^0_+ \tilde{b}^0\nonumber\\
\delta h=- \frac{g}{2}c^a\tilde{\varphi}^a_-- \tilde{b}^0\nonumber\\
\delta \tilde{\varphi}_+^a=\frac{g}{2} \epsilon^{abc}\tilde{\varphi}_+^bc^c+
 \frac{g}{2} \tilde{\varphi}_+^0 c^a- \frac{g^2}{4m}\tilde{\varphi}_+^a \tilde{\varphi}^b_- c^b- \frac{g}{2m} \tilde{\varphi}^a_+ \tilde{b}^0 \nonumber\\
\delta \tilde{\varphi}^a_-=mc^a+\frac{g}{2} \epsilon^{abc}\tilde{\varphi}_-^b c^c+\frac{g^2}{4m}\tilde{\varphi}_-^a \tilde{\varphi}^b_- c^b- \tilde{b}^a + \frac{g}{2m} \tilde{\varphi}^a_- \tilde{b}^0\nonumber\\
\delta \tilde{b}^a=-\frac{g}{2} \epsilon^{abc}\tilde{b}^bc^c-\frac{g}{2} \tilde{b}^0c^a+\frac{g^2}{4m} \tilde{\varphi}_-^b
\tilde{b}^ac^b- \frac{g}{2m} \tilde{b}^a \tilde{b}^0\nonumber\\
\delta \tilde{e}^a=-\frac{g}{2} \epsilon^{abc}\tilde{e}^bc^c-\frac{g}{2}\tilde{e}^0c^a+ \frac{g^2}{4m} \tilde{\varphi}_-^b \tilde{e}^ac^b+ \tilde{\varphi}^a_+ + \frac{g}{2m} \tilde{e}^a \tilde{b}^0\nonumber\\
\delta \tilde{b}^0=+\frac{g}{2}\tilde{b}^a c^a- \frac{g^2}{4m}\tilde{\varphi}_-^b
\tilde{b}^0c^b\nonumber\\
\delta \tilde{e}^0= +\frac{g}{2} \tilde{e}^a c^a+ \frac{g^2}{4m}\tilde{\varphi}_-^b \tilde{e}^0c^b+ \tilde{\varphi}^0_+ +\frac{g}{2m} \tilde{e}^0 \tilde{b}^0 \label{10a}
\end{eqnarray}
 It allows to use instead of the canonical gauge fixing
 eq.(\ref{7a}) the following gauge fixing
\begin{equation}
s^1 \int d^4x \bar{c}^a \tilde{\varphi}^a_-= \int d^4x(\lambda^a \tilde{\varphi}^a_-- \bar{c}^a(M^{ab}c^b- \tilde{b}^a) ) \, ,
\label{11a}
\end{equation}

where $s^1$ is the nilpotent operator defined by the eqs.(\ref{9}) and
(\ref{10a}).

The scattering matrix may be alternatively presented either in the canonical form
\begin{equation}
S=\int \exp\{i\int [\tilde L+ \lambda^a \tilde{\varphi}^a_-- \bar{c}^a M^{ab} c^b]d^4x\}d \mu \label{12a}
\end{equation}
 or in the form
\begin{equation}
S=\int \exp\{i\int [\tilde L+ \lambda^a \tilde{\varphi}^a_-- \bar{c}^a M^{ab} c^b+ \bar{c}^a \tilde b^a]d^4x\}d \mu
\label{13a}
\end{equation}
where $L$ is the Lagrangian (\ref{6c}).

The integral (\ref{13a}) may be transformed to the form (\ref{12a}) by the change of variables
\begin{equation}
c^a \rightarrow c^a+(M^{-1} \tilde{b})^a \label{14a.bis}
\end{equation}
Note that in the gauge $\tilde{\varphi}^a_-=0$ the operator $M^{ab}$ is equal to $\delta^{ab}m$.

As the variables $ \bar{c}, c$ are ultralocal, this transformation does not change the value of
the integral. Hence for studying the structure of the counterterms necessary for perturbative
renormalization of these integrals one may use the symmetries of both integrals (\ref{12a}) and (\ref{13a}).

Performing the integration over $\bar{c},c$ in the eq.(\ref{13a})
we obtain in the exponent the
effective action
which is invariant with respect to the transformation
 obtained from
the transformations eq.(\ref{10a}) by substituting  $c^a=(M^{-1}\tilde{b})^a$:
\begin{eqnarray}
\delta A_{\mu}^a=\frac{1}{m}(D_{\mu}\tilde{b})^a\nonumber\\
\delta\tilde{\varphi}_-^a=0\nonumber\\
\delta h= -\tilde{b}^0\nonumber\\
\delta \tilde{\varphi}_+^a= \frac{g}{2m} \tilde{\varphi}^0_+ \tilde{b}^a+\frac{g}{2m} \varepsilon^{abc} \tilde{\varphi}_+^b \tilde{b}^c-\frac{g}{2m} \tilde{\varphi}^a_+\tilde{b}^0\nonumber\\
\delta\tilde{\varphi}_+^0=-\frac{g}{2m}( \tilde{\varphi}^a_+ \tilde{b}^a+ \tilde{\varphi}^0_+ \tilde{b}^0)\nonumber\\
\delta \tilde{e}^a= \frac{g}{2m}( \tilde{e}^a \tilde{b}^0 - \tilde{e}^0 \tilde{b}^a- \varepsilon^{abc} \tilde{e}^b \tilde{b}^c)+ \tilde{\varphi}^a_+\nonumber\\
\delta \tilde{b}^0=0\nonumber\\
\delta \tilde{e}^0=\frac{g}{2m}( \tilde{e}^a \tilde{b}^a+ \tilde{e}^0 \tilde{b}^0)+ \tilde{\varphi}^0_+ ~~~~~
\delta \tilde b^a=- \frac{g}{2m} \varepsilon^{abc} \tilde{b}^b \tilde{b}^c
\label{14a}
\end{eqnarray}
One sees that these transformations do not change the field
$\tilde{\varphi}^a_-$  , hence the effective
action in the gauge $\tilde{\varphi}^a_-=0$   is invariant under these transformations.

The effective action in the gauge $\tilde{\varphi}^a_-=0$  is given by the eq.(\ref{6c}) with all terms $\sim \tilde{\varphi}^a_-$ omitted.
This action obviously is not invariant with respect to the global $SU(2)$ rotations of all
variables, but it is still invariant with respect to the $SU(2)$ transformations which do not
change the fields $\tilde{\varphi}^0_+,\tilde{\varphi}^a_-, h$. This symmetry will be very helpful for analysis
of possible counterterms.

The quadratic part of the effective action is
 \begin{eqnarray}
A_0=\int d^4x[-\frac{1}{4}(\partial_{\mu}A_{\nu}^a-\partial_{\nu}A_{\mu}^a)^2+
\partial_{\mu}h \partial_{\mu} \tilde{\varphi}^0_+\nonumber\\
_+m \tilde{\varphi}^a_+\partial_{\mu}A_{\mu}^a -\tilde{b}^0\partial^2 \tilde{e}^0-\tilde{b}^a
\partial^2 \tilde{e}^a]
\label{18a}
\end{eqnarray}

The nonvanishing propagators are:
\begin{eqnarray}
\Delta(A^a_{\mu}A^b_{\nu})=\frac{-i\delta^{ab}}{p^2}T_{\mu\nu}, \quad
\Delta(A^a_{\mu}\tilde{\varphi}^b_+)=-\delta^{ab} \frac{p_{\mu}}{mp^2}\nonumber\\
\Delta(\tilde{\varphi}^0_+h)=\frac{i}{p^2}, \quad \Delta(\tilde{b}^0 \tilde{e}^0)=\frac{i}{p^2}, \quad
\Delta(\tilde{b}^a \tilde{e}^b)=\frac{i\delta^{ab}}{p^2} \label{19a}
\end{eqnarray}
One sees that the complete asymptotic space of our model includes a number of unphysical
excitations corresponding to the fields $\tilde{\varphi}^{\pm}, \tilde{b}, \tilde{e}, h$   and longitudinal and temporal
components of the Yang-Mills field. Note however that the Faddeev-Popov ghosts $\bar{c}^a, c^a$
are ultralocal and do not enter the asymptotic states. The unitarity of the model in the space
including only three dimensionally transversal components of the Yang-Mills field is provided by
the symmetry (\ref{14a}). The corresponding symmetry for asymptotic fields generates the
nilpotent conserved charge $Q^0$, which acts on the asymptotic fields as follows
\begin{eqnarray}
Q^0A^a_{\mu}=\frac{1}{m}\partial_{\mu} \tilde{b}^a \nonumber\\
Q^0 \tilde{e}^a= \tilde{\varphi}^a_+ \label{20a}
\end{eqnarray}

One sees that the field  $\tilde{e}^a$  may be identified with the antighost field of the Yang-Mills
theory. It is mapped into   $\tilde{\varphi}^a_+$   which plays the role of the Nakanishi-Lautrup field
implementing the Lorentz gauge condition. The field
$\tilde{b}^a$ plays the role of the Yang-Mills ghost.
Therefore the transformations (\ref{14a})  asymptotically   coincide with the usual BRST transformations which
provide the decoupling of the fields    $\tilde{\varphi}^a_+, \tilde{e}^a, \tilde{b}^a$  and unphysical components of the
Yang-Mills field. Moreover
\begin{eqnarray}
Q^0 \tilde{e}^0=\tilde{\varphi}^0_+\nonumber\\
Q^0 \tilde{\varphi}^0_+=0\nonumber\\
Q^0 h= -\tilde b^0\nonumber\\
Q^0 \tilde{b}^0=0 \label{21a}
\end{eqnarray}

which guarantees the decoupling of   $\tilde{e}^0, \tilde{b}^0, \tilde{\varphi}^0_+, h$  from the physical subspace. The
ultralocal fields $\lambda^a, \bar{c}^a, c^a$ do not contribute to the physical asymptotic
states.

Therefore formally the model described above has the same
spectrum of observables as the usual Yang-Mills model,
and one can show that the correlators of the gauge invariant
operators also coincide. However our discussion so far was formal
as we did not take into account the necessity of renormalization and
dealt with ultraviolet divergent integrals.

The problem of renormalization in the gauge $\tilde \varphi^a_-=0$ appears to be quite nontrivial.
Renormalization of our model requires not only changing the values of the
parameters in the classical Lagrangian, but also a
redefinition of the fields.
So our model
gives an example of a "General Gauge Theory" considered in details
by B.L.Voronov and I.V.Tyutin~\cite{VT}.

\section{Renormalization} \label{sec.3}


 The divergency index of an arbitrary diagram is equal to
\begin{equation}
n = 4-2L_{\varphi^0_+}-2L_{\varphi^a_+}-L_A-L_e-L_b-L_h \label{22a}
\end{equation}
where $L_\Phi$ is the number of the external
lines of the field $\Phi$.
For the diagrams with two or more external lines $n \leq 2$. Any diagram with more than four external lines is convergent, and the theory is manifestly renormalizable.

It is worth to notice that the transition to the variables (\ref{6a}), performed to get rid off the factor $m+ \frac{g}{2} \varphi^0_-$, which may lead to ambiguity, is needed also for a manifest renormalizability of the theory. If we quantize the theory in the original variables, the expression for the divergency index would be given by the eq.(\ref{22a}) without the last term. In that case the divergency index does not depend on the number of external lines of the field $\varphi^0_-$ ($h$-lines in the new variables). As a result there are divergent diagrams with arbitrary number of external $\varphi^0_-$ lines. The transformation (\ref{6a}) cures simultaneously two diseases: it eliminates a possible source of ambiguity and makes the model manifestly renormalizable.

\medskip
However the action
\begin{eqnarray}
A_{ef} = \int d^4x \tilde L \, ,
\label{eff.act.0}
\end{eqnarray}
where $\tilde L$ is the Lagrangian in eq.(\ref{6c}), is not the most general
classical action which is invariant under the transformation
(\ref{14a}) and the global $SU(2)$ transformations
of the fields
$A^a_\mu, \tilde \varphi^a_+, \tilde e,\tilde b$.
The following combination also respects the invariance
under the transformation
(\ref{6c}) as well as the residual $SU(2)$ invariance:
\begin{eqnarray}
{\cal G} &= & \int d^4x \, \Big [
(\tilde \varphi^0_+)^2 + (\tilde \varphi^a_+)^2
+ \frac{g}{m} \tilde \varphi^0_+ (\tilde e^0 \tilde b^0 + \tilde e^a \tilde b^a)
+ \frac{g}{m} \tilde \varphi^a_+ (
\tilde e^a \tilde b^0
-\tilde e^0 \tilde b^a
- \varepsilon^{abc} \tilde e^b \tilde b^c ) \nonumber \\
&&~~~~~~~ -\frac{g^2}{2m^2}
\Big ( - (2\tilde e^0 \tilde b^0 + \tilde e^a \tilde b^a) \tilde e^b \tilde b^b
+ \varepsilon^{abc} \tilde e^0 \tilde b^b \tilde e^c \tilde b^a
- \varepsilon^{abc} \tilde e^b \tilde b^0 \tilde e^c \tilde b^a
\Big )
\Big ] \, .
\label{comb.1}
\end{eqnarray}
Moreover it does not violate
power-counting renormalizability and
therefore it  can be introduced
in the action (\ref{eff.act.0}).
${\cal G}$ can be
made gauge-invariant by adding terms
proportional to $\tilde \varphi^-_a$.
This can be easily done by performing
in eq.(\ref{comb.1}) the substitutions
\begin{eqnarray}
&& \tilde \varphi^0_+ \rightarrow
     \tilde \varphi^0_+ + \frac{g}{2m} \tilde \varphi^a_- \tilde \varphi^a_+ \, , ~~~~
     \tilde \varphi^a_+ \rightarrow
     \tilde \varphi^a_+ - \frac{g}{2m} \tilde \varphi^a_- \tilde \varphi^0_+ + \frac{g}{2m} \varepsilon^{abc} \tilde \varphi^b_- \tilde \varphi^c_+ \, , \nonumber \\
&& \tilde e^0 \rightarrow
     \tilde e^0 + \frac{g}{2m} \tilde \varphi^a_- \tilde e^a \, , ~~~~~~~~
     \tilde e^a \rightarrow
     \tilde e^a - \frac{g}{2m} \tilde \varphi^a_- \tilde e^0 + \frac{g}{2m} \varepsilon^{abc} \tilde \varphi^b_- \tilde e^c \, , \nonumber \\
&& \tilde b^0 \rightarrow
     \tilde b^0 + \frac{g}{2m} \tilde \varphi^a_- \tilde b^a \, , ~~~~~~~~
     \tilde b^a \rightarrow
     \tilde b^a - \frac{g}{2m} \tilde \varphi^a_- b^0 + \frac{g}{2m} \varepsilon^{abc} \tilde \varphi^b_- \tilde b^c \, .
\label{new.gauge.inv.var}
\end{eqnarray}
The expressions in the R.H.S. of  eqs.(\ref{new.gauge.inv.var})  are invariant
w.r.t. the gauge transformation eq.(\ref{6d}), as can be checked
by direct computation.
Gauge-invariance fixes uniquely the dependence
on $\tilde \varphi^a_-$.


The new action is
\begin{eqnarray}
A'_{ef} & = & A_{ef} + \frac{m^2}{2} \alpha {\cal G} \, .
\label{comb.2}
\end{eqnarray}
The prefactor $m^2$ in front
of ${\cal G}$ has been inserted
for dimensional reasons and $\alpha$ plays the role of the gauge-fixing parameter.

By keeping the full dependence
on $\tilde \varphi^a_-$
one explicitly finds
\begin{eqnarray}
A'_{ef} & = &
\int d^4x \Big \{ -\frac{1}{4}F_{\mu\nu}^a F_{\mu\nu}^a +
\partial_\mu h \partial^\mu \tilde \varphi^0_+
+ m \partial A^a \tilde \varphi_+^a
+ \partial_\mu \tilde \varphi_-^a \partial^\mu \tilde \varphi_+^a
- \frac{g}{2m} \partial_\mu h \partial^\mu h \tilde \varphi_+^0
\nonumber \\
&&~~~~~~~~ + g \partial_\mu h A^a_\mu\tilde \varphi_+^a
+ \frac{gm}{2} A_{\mu}^2 \tilde \varphi^0_+
+ \frac{g}{2m} \partial_\mu h
(\partial^\mu \tilde \varphi^a_+ \tilde \varphi^a_-
- \partial^\mu \tilde \varphi^a_- \tilde \varphi^a_+)
\nonumber \\
&& ~~~~~~~~ + \frac{g}{2} A^a_\mu
(\partial^\mu \tilde \varphi^0_+ \tilde \varphi^a_- -
 \tilde \varphi^0_+ \partial^\mu \tilde \varphi^a_-)
 + \frac{g}{2} \varepsilon^{abc} A^a_\mu
 (\tilde \varphi^b_- \partial^\mu \tilde \varphi^c_+
 - \partial^\mu \tilde \varphi^b_- \tilde \varphi^c_+)
 \nonumber \\
&& ~~~~~~~~ + \frac{g^2}{4}
A_{\mu}^2 \tilde \varphi_-^b \tilde \varphi_+^b
- \frac{g^2}{4m^2} \partial_\mu h \partial^\mu h
\tilde \varphi^a_- \tilde \varphi^a_+
- \frac{g^2}{2m} \partial^\mu h \tilde \varphi^0_+ A^a_\mu
\tilde \varphi^a_-
\nonumber \\
&& ~~~~~~~~ +
\frac{g^2}{2m} \partial_\mu h ~
\varepsilon^{abc} \tilde \varphi^a_+ \tilde \varphi^b_- A^{c\mu}
\nonumber
-[((D_{\mu} \tilde{b})^*+ \frac{g}{2m} \tilde{b}^*\partial_{\mu}h)((D_{\mu} \tilde{e})- \frac{g}{2m} \tilde{e}\partial_{\mu}h)+h.c.]\nonumber \\
&& ~~~~~~~~
+ \frac{m^2 \alpha}{2}
\Big [ \Big ( \tilde \varphi^0_+ + \frac{g}{2m}
\tilde \varphi^a_- \tilde \varphi^a_+
+ \frac{g}{2m} (\tilde e^0 \tilde b^0 +
\tilde e^a \tilde b^a) \Big )^2 \nonumber \\
&& ~~~~~~~~~
+ \Big ( \tilde \varphi^a_+ - \frac{g}{2m} \tilde \varphi^0_+ \tilde \varphi^a_- - \frac{g}{2m} \varepsilon^{abc} \tilde \varphi^b_-\tilde \varphi^c_+
+ \frac{g}{2m}
(
\tilde e^a \tilde b^0
-\tilde e^0 \tilde b^a
- \varepsilon^{abc} \tilde e^b \tilde b^c )
\Big )^2 \Big ]
\Big \} \nonumber \\
\label{full.form}
\end{eqnarray}
which is manifestly polynomial.



The quadratic part of $A'_{ef}$
in the gauge $\tilde \varphi^-_a=0$
reads
\begin{eqnarray}
A'_{ef,0}& = &\int d^4x[-\frac{1}{4}(\partial_{\mu}A_{\nu}^a-\partial_{\nu}A_{\mu}^a)^2+
\frac{\alpha}{2} m^2 (\tilde \varphi^0_+)^2 - \tilde \varphi_0^+ \square h
\nonumber\\
&& ~~~~~~~+m \tilde \varphi^a_+\partial_{\mu}A_{\mu}^a+ \frac{\alpha}{2} m^2 (\tilde \varphi^a_+)^2
-\tilde b^0\partial^2\tilde e^0-\tilde b^a
\partial^2\tilde e^a] \,.
\label{prop.1}
\end{eqnarray}
The nonvanishing propagators are
\begin{eqnarray}
&&
\!\!\!\!\!\!\!\!
\Delta(A^a_\mu A^b_\nu) =
\frac{-i \delta^{ab}}{p^2} T_{\mu\nu} + i \delta^{ab} \frac{\alpha}{p^2} L_{\mu\nu} \, , ~~~~~
\Delta(A^a_\mu \tilde \varphi_+^b) = -\delta^{ab} \frac{p_\mu}{mp^2} \, ,  \nonumber \\
&&
\!\!\!\!\!\!\!\!
\Delta(\tilde \varphi^0_+ h) = \frac{i}{p^2} \, , ~~~~
\Delta(hh) = -\frac{i \alpha m^2}{p^4} \, ,
~~~~
\Delta(\tilde b^0 \tilde e^0) = \frac{i}{p^2} \, , ~~~~ \Delta(\tilde b^a\tilde e^b) = \frac{i\delta^{ab}}{p^2} \, .
\label{prop.2}
\end{eqnarray}
The gauge field is quantized in the $\alpha$-gauge, while a dipole
arises for $h$. However, we notice that the dependence on $\varphi^+_0,h$ in eq.(\ref{prop.1})
can be recast as
\begin{eqnarray}
\int d^4x \, \Big [ \frac{\alpha}{2} \Big ( \tilde \varphi^0_+ - \frac{1}{\alpha} \square h \Big )^2
+ \frac{1}{2\alpha} \partial_\mu h \square \partial_\mu h \Big ] \, .
\label{prop.3}
\end{eqnarray}
The combination $\tilde \varphi^0_+ - \frac{1}{\alpha} \square h$ is ultralocal.
Moreover the decoupling of $\partial_\mu h$ from the physical states is
guaranteed by the symmetry (\ref{21a}) since at the asymptotic level
\begin{eqnarray}
&& Q h = - \tilde b_0 \, , \nonumber\\
&& Q \tilde b_0 = 0 \, .
\label{prop.4}
\end{eqnarray}
So the nonrenormalized action (\ref{comb.2})
preserves the unitarity in the subspace including
only three dimensionally transversal components
of the Yang-Mills field.

\section{The structure of counterterms}\label{sec.4}

In the previous Section we showed that the action
(\ref{comb.2}) describes a renormalizable theory
which formally is unitary in the physical subspace.
The crucial role in the proof of the unitarity was played
by the symmetry of the theory with respect to
the transformations (\ref{14a}).

Now we are going to prove that renormalization does not
violate this property. The counterterms needed to remove
all ultraviolet divergences only change the values
of the parameters entering the action
(\ref{comb.2}) (modulo field redefinitions) and the
transformations (\ref{14a}), but preserve the symmetry
of the renormalized theory. Moreover we shall
demonstrate that the renormalized action
in the gauge $\tilde\varphi^a_-=0$  may be obtained
by imposing the gauge condition on the invariant
classical action (assuming of course that
some gauge invariant intermediate regularization
is introduced).
Using this fact one can easily prove
gauge independence of the observables.
In particular in perturbation theory, when the
Gribov ambiguity is absent, the correlation
functions of gauge invariant operators
constructed
from the Yang-Mills field
calculated
in the ambiguity free gauge and in the
Lorentz gauge $\partial_\mu A_\mu=0$
coincide.

In order to study the counterterms in the
gauge $\tilde\varphi^a_-=0$ let us introduce a new action $\G^{(0)}$, including apart from the classical
action $A'_{ef}$ in eq.(\ref{comb.2}) also the variation of the fields
$\Phi$, coupled to the external sources $\Phi^*$, which are
usually called ''antifields'' \cite{ZJ}, i.e. we set
\begin{eqnarray}
\G^{(0)} = A'_{ef} + \sum_\Phi \int d^4x \, \Phi^* \delta \Phi \, .
\label{vertex.functional.tl}
\end{eqnarray}
where $\delta$ is defined in eq.(\ref{14a}).
Then the invariance under the symmetry (\ref{14a}) is translated into the
following functional identity for the 1-PI generating functional $\G$
\begin{eqnarray}
{\cal S}(\G) = \int d^4x \, \sum_\Phi \frac{\delta \G}{\delta \Phi^*(x)}
\frac{\delta \G}{\delta \Phi(x)}
= 0 \, .
\label{sc.1}
\end{eqnarray}
$\G$ is developed in the number of loops, i.e. $\G = \sum_{j=0}^\infty \hbar^j \G^{(j)}$.

Assuming that some invariant regularization is introduced,
eq.(\ref{sc.1}) holds if the effective action $\hat \G$
(tree-level plus counterterms) fulfills
\begin{eqnarray}
{\cal S}(\hat \G) = 0 \, .
\label{sc.2}
\end{eqnarray}
Moreover we will also require invariance
 under the residual global $SU(2)$ symmetry.

The most general solution of the eq.(\ref{sc.2}) compatible
with the dimensional bounds and the residual $SU(2)$ invariance may
be written as follows. One should make the shift of the parameters
$g,m, \alpha$,  which enter the classical action $A_{ef}'$
(\ref{comb.2}) and redefine the fields preserving the
ultraviolet counting (eq.(\ref{22a})):
\begin{eqnarray}
g' = Z_g g \, , ~~~ m' = Z_m m \, , ~~~ \alpha' = \frac{Z_\alpha}{Z_m^2} \alpha \, ,
\label{sc.2.1}
\end{eqnarray}
\begin{eqnarray}
&& \tilde e' = Z_1 \tilde e \, , ~~~~  \tilde b' = Z_m \tilde b \, , ~~~~  A^{a'}_\mu = Z_2 A^a_\mu \, ,
~~~~ h' = Z_m Z_3 h \, , \nonumber \\
&& \tilde \varphi^{a'}_+ =
Z_4 \tilde \varphi^a_+ +
Z_5 \partial A^a + Z_6 \frac{1}{m} \partial_\mu h A^{a\mu}
+ Z_7 (\tilde e^0 \tilde b^a - \tilde e^a \tilde b^0 -
\varepsilon^{abc} \tilde e^b \tilde b^c ) \, , \nonumber \\
&& \tilde \varphi^{0'}_+ =
Z_8 \tilde \varphi^0_+ + Z_9 \frac{1}{m} \square h +
Z_{10} \frac{1}{m^2} \partial_\mu h \partial^\mu h + Z_{11} A^2
+Z_{12} (\tilde e^0 \tilde b^0 + \tilde e^a \tilde b^a) \, .
\label{sc.3}
\end{eqnarray}
Since a multiplicative redefinition of $\tilde b$ can always be
compensated in $A'_{ef}$ by a redefinition of $\tilde e$,  we do not
rescale $\tilde b$ apart from a factor $Z_m$, which is introduced
for convenience in such a way that $Z_m$ multiplies in $\hat \G$ the
global $SU(2)$- and $\delta$-invariant combination containing the
kinetic term for $\tilde e, \tilde b$.
Note that  to satisfy
eq.(\ref{sc.2}) the redefinition of the fields must be supplemented
by the corresponding redefinition of the antifields.

For that purpose one may notice that the functional
identity (\ref{sc.1}) can be formulated by means of the
following bracket \cite{Troost:1989cu}
\begin{eqnarray}
(X,Y) = \int d^4x \sum_\Phi (-1)^{\epsilon(\Phi)\epsilon(X)}
\Big (
\frac{\delta X}{\delta \Phi} \frac{\delta Y}{\delta \Phi^*}
- (-1)^{\epsilon(X)+1} \frac{\delta X}{\delta \Phi^*}\frac{\delta Y}{\delta \Phi} \Big )
\label{sc.4}
\end{eqnarray}
where $\epsilon$ denotes the statistics
($1$ for fermions, $0$ for bosons). I.e. one has
\begin{eqnarray}
{\cal S}(\G) = \frac{1}{2} (\G,\G) = 0 \, .
\label{bracket.fi}
\end{eqnarray}
Under eq.(\ref{sc.4}) the fields and the antifields are paired via the
fundamental brackets
\begin{eqnarray}
(\Phi_i,  \Phi^*_j ) =(-1)^{\epsilon(\Phi_j)} \delta_{ij} \, .
\label{bracket.fund}
\end{eqnarray}
Notice that our conventions on the antifields differs from the one
of \cite{Troost:1989cu} by the redefinition $\Phi^* \rightarrow
(-1)^{\epsilon(\Phi)} \Phi^*$, whence the sign factor in the r.h.s.
of the above equation.

A redefinition of the fields and the antifields preserving eq.(\ref{bracket.fund})
is called a canonical transformation (w.r.t. the bracket (\ref{sc.4})).
It automatically preserves the bracket between any two functionals $X,Y$ and
therefore also the functional identity (\ref{sc.2}).

A convenient way to complete the field redefinition (\ref{sc.3}) to
a finite canonical transformation is to derive the latter via a
suitable generating functional~\cite{Troost:1989cu}. In the present
case this is given by  $G = \int d^4x \sum_{\Phi'}
(-1)^{\epsilon(\Phi')} \Phi^{*'} \Phi'(\Phi)$ and the field and
antifield transformations are obtained
 by solving the equations
\begin{eqnarray}
\Phi^* = (-1)^{\epsilon(\Phi)} \frac{\delta G}{\delta \Phi} \, , ~~~
\Phi' = (-1)^{\epsilon(\Phi')} \frac{\delta G}{\delta \Phi^{*'}} \, .
\label{sc.fin}
\end{eqnarray}
By construction $G$ generates the field redefinition (\ref{sc.3}), while
the explicit expressions for the antifield redefinitions are presented in Appendix ~\ref{app.A}.

Consequently the functional
\begin{eqnarray}
\hat \G[ g',m',\alpha'; \Phi', \Phi^{*'}] =
 \G^{(0)}[Z_g g, Z_m m, Z_\alpha/Z_m^2 \alpha; \Phi(\Phi'),\Phi^{*}(\Phi',\Phi^{*'})]
\label{sc.5}
\end{eqnarray}
is a solution to eq.(\ref{sc.2}). One can verify it by
explicit calculations.

It remains to be shown that all
the divergences can be recursively removed by a suitable choice of
the parameters  $Z_g,Z_m,Z_\alpha$ and by
changing the field renormalization constants
$Z_j$, $j=1,\dots,12$.
This is done in Appendix~\ref{app.B}.

As announced, the renormalized effective action $\hat
\G$ is finally obtained by a shift in the parameters of the
classical action (modulo field redefinitions).

\section{Independence of observables on the gauge and comparison with the
standard Yang-Mills theory}\label{sec.5}

In the previous sections we proved that all the
ultraviolet
divergencies in the gauge invariant theory determined by the action
(\ref{1}) plus the additional term (\ref{comb.1}), quantized in the gauge
$\tilde\varphi^a_-=0$, may be removed by the renormalization of the
parameters entering the classical action and the local redefinition
of the fields. It was shown that the resulting (infrared
regularized) scattering matrix is unitary in the subspace including
only three dimensionally transversal components of the Yang-Mills
field.

Now we want to show that the scattering matrix, obtained in this
way, as well as other gauge invariant correlators depending only on
the Yang-Mills field
 in the
framework of perturbation theory
may be transformed to
the normally used differential gauges.
In particular they can be calculated in the Lorentz
gauge $\partial_{\mu}A_{\mu}=0$. The values of these quantities also
coincide with the corresponding values in the standard Yang-Mills
theory.

The renormalized theory in the gauge $\tilde\varphi^a_-=0$ is described by
the effective action (\ref{sc.5}) with the proper chosen constants
$Z_g, Z_m, Z_{\alpha}$ and $Z_j$. This renormalized action was
obtained from the gauge invariant classical action by
imposing the condition $\tilde\varphi^a_-=0$ and redefining the fields and
the parameters. As all the field redefinitions were local, in the
path integral formulation they lead to the appearance of local
jacobians, which in regularizations like dimensional one are
trivial. Having in mind this kind of invariant regularization we can
perform the inverse redefinition of the fields, resulting in the
original classical action depending on the renormalized parameters.
Note that the redefinition of the fields used above did not include
the field $\tilde\varphi^a_-$.

Therefore the scattering matrix may be presented as
the path integral
\begin{equation}
S=\int \exp\{i\int[L_{g.i.}+ \lambda^a \varphi^a_-]dx\}
\det(M_{ab})d\mu \label{56}
\end{equation}
where the local Jacobian $\det(M_{ab})$ replaces the usual
Faddeev-Popov determinant and integration goes over all fields
present in the Lagrangian. The boundary conditions for the three
dimensionally transversal components of the Yang-Mills field are
determined by the corresponding asymptotic fields and
for all other fields we may use the vacuum (radiation) conditions,
as the scattering matrix is unitary in the space including only
three dimensionally transversal components of the Yang-Mills field.
The gauge invariant
Lagrangian $L_{g.i.}$ depends on the renormalized parameters
\begin{equation}
g_R=Z_g g; \quad m_R=Z_m m; \quad \alpha_R= \alpha
\frac{Z_{\alpha}}{Z_m^2} \label{57}
\end{equation}
As usual the jacobian $\det(M_{ab})$ may be presented as follows
\begin{equation}
(\det(M_{ab})^{-1})_{\varphi^a_-=0}=
\int\delta((\tilde\varphi^\Omega)^a_-)d\Omega \label{58}
\end{equation}

\medskip

Multiplying the integral (\ref{56}) by $"1"$
\begin{equation}
1=\Delta_L \int \delta(\partial_{\mu}A_{\mu}^\Omega)d\Omega
\label{59}
\end{equation}
and changing the variables
\begin{equation}
\Phi^\Omega=\Phi' \label{60}
\end{equation}
we obtain the expression for the scattering matrix in the Lorentz
gauge:
\begin{equation}
S =\int \exp\{i\int[L_{g.i.}(x)+
\lambda^a(x)\partial_{\mu}A^a_{\mu}(x)+\partial_{\mu}\bar{c}^a[D_{\mu}c]^a]dx\}d\mu
\label{61}
\end{equation}
where $ \bar{c}, c$ are the usual Faddeev-Popov ghosts.

These reasonings show that the scattering matrix as well as the
gauge invariant correlators depending only on 
the Yang-Mills field computed in the ambiguity free gauge
$\tilde\varphi^a_-=0$ coincide with the corresponding objects in the
Lorentz gauge. Strictly speaking this transition is legitimate only
in the framework of the perturbation theory as beyond the
perturbation theory the Faddeev-Popov determinant may have zeroes.

The last thing we wish to show is the equality of the scattering
matrix and the gauge invariant correlators calculated on the basis
of the action (\ref{1}) plus the additional terms (\ref{comb.1}) in the
Lorentz gauge $\partial_{\mu}A_{\mu}=0$ to the corresponding objects
in the standard Yang-Mills theory.

To do that we again make the inverse transformation of the fields,
writing the eqs.~(\ref{1}) and (\ref{comb.1}) in terms of the original variables .
The eq.(\ref{1}) has the same form as before. The only difference is the
change of the charge by the renormalized charge.

The  expression for ${\cal G}$ may be written
in terms of the original fields as follows
\begin{eqnarray}
{\cal G} & = &  \int d^4x \, \Big \{
\Big [ \frac{g}{2m} \Big ( (\varphi^*_- + \hat m^*) \varphi_+ +
\varphi_+^* (\varphi_- + \hat m) \Big )  +
\frac{g}{2m} (b^* e + e^* b) \Big ]^2 \nonumber \\
&&~~~~~~ + \Big [
\frac{g}{2m} \Big (
- i ( \varphi^*_- + \hat m^*) \tau^a \varphi_+ + i
               \varphi^*_+ \tau^a (\varphi_- + \hat m)  \Big )
               + \frac{ig}{2m}  (b^* \tau^a e - e^* \tau^a b)
               \Big ]^2
\Big \} \nonumber \\
\label{n.3.new}
\end{eqnarray}
Introducing the local field $\mu^A$, $A=0,1,2,3$ one can represent the exponent
$\exp \{i\frac{m^2 \alpha}{2}{\cal G} \}$ as follows:
\begin{eqnarray}
&&
\!\!\!\!\!\!\!\!\!\!\!\!\!\!\!\!\!\!\!
 \exp \{i \frac{m^2 \alpha}{2} {\cal G}\} = \int \exp\{ -i
m^2 \alpha
\int \Big [  -\frac{g}{2m} \Big ( (\varphi^*_- + \hat m^*) \varphi_+ +
\varphi_+^* (\varphi_- + \hat m) +
(b^* e + e^* b)  \Big ) \mu^0 + \frac{(\mu^0)^2}{2}
\nonumber\\
&& ~~~~~~~~
-
\frac{g}{2m} \Big (
- i ( \varphi^*_- + \hat m^*) \tau^a \varphi_+ + i
               \varphi^*_+ \tau^a (\varphi_- + \hat m)
               + i (b^* \tau^a e - e^* \tau^a b)
               \Big ) \mu^a +
\frac{(\mu^a)^2}{2}
]dx\}d\mu
\nonumber \\
\label{63}
\end{eqnarray}

Changing the variables as in the paper~\cite{Sl1}
\begin{eqnarray}
\varphi(x)=\varphi'(x)+g^{-1} \int D^{-2}(x,y)(D^2
\hat{m}(y))dy\nonumber\\
 \chi(x)=\chi'(x)-g^{-1} \int D^{-2}(x,y)(D^2
\hat{m}(y))dy
\label{64}
\end{eqnarray}
and integrating over auxilliary fields $\varphi, \chi, b, e$ we
obtain the determinants which cancel each other and finally we are
left with the expression which coincides with the standard
Yang-Mills theory.

 Note that in the equation (\ref{64})we are
allowed to perform the integration by parts as the corresponding
expressions are multiplied by the functions decreasing at infinity.
We also may integrate explicitely over the auxilliary fields with
the vacuum boundary conditions, as above
 we proved the unitarity of
the scattering matrix in the subspace including only three
dimensionally transversal components of the Yang-Mills field.

\section{Discussion}\label{sec.6}

In this paper we showed that the Yang-Mills theory allows a renormalizable formulation free of
the Gribov ambiguity. It provides strong arguments in favour of the point of view according to
which this ambiguity is an artefact of the quantization procedure and cannot produce some
physical effects. From the technical point of view the model considered in this paper gives an
interesting example of a  nontrivial renormalizable theory whose renormalization requires
nonmultiplicative field redefinition.

Of course a rigorous comparison of different formulations is possible only in the framework of
the perturbation theory where the ambiguity is absent and it is not excluded that beyond the
perturbation theory our formulation and the standard one describe different theories. However
such a possibility seems to be rather unlikely. It is worth to mention that the studies of
gluodynamics beyond the perturbation theory carried out now both by semi analytic methods,
mainly based on the Schwinger-Dyson equations \cite{Betal,
ABP, FMP}, and by computer simulations
\cite{CM,BIMS}, give controversial results concerning the infrared behaviour of the propagators. It
would be interesting to carry out similar investigations in the present formulation.

\section*{Acknowledgments}

This work was started when one of the present authors (A.A.S.) was
visiting the University of Milan. A.A.S. thanks R.~Ferrari for hospitality
and Cariplo Foundation for the support. The work of A.A.S. was partially
supported by the RBRF grant
09-01-12150-{\cyr ofi\_m}
and RAS program "Nonlinear dynamics."

\appendix
\section{Antifield Transformations}\label{app.A}

We collect here the explicit form of the antifield redefinitions induced by the canonical
transformation (\ref{sc.fin}):
\begin{eqnarray}
&&
{\tilde e}^{0*} = Z_1 {\tilde e}^{0*'}
                            + Z_7 {\tilde \varphi}^{a*'}_+ {\tilde b}^a
                            + Z_{12} {\tilde \varphi}^{0*'}_+ {\tilde b}^0 \, , \nonumber \\
&&
 {\tilde e}^{a*} = Z_1 {\tilde e}^{a*'} -
                 Z_7   {\tilde \varphi}^{a*'}_+ {\tilde b}^0 +
                 Z_7 \varepsilon^{abc}  {\tilde \varphi}^{b*'}_+ {\tilde b}^c
                 + Z_{12} {\tilde \varphi}^{0*'}_+ {\tilde b}^a \, , \nonumber \\
&&
{\tilde A}^{a*}_\mu = Z_2 A^{a*'}_\mu
                 - Z_5 \partial^\mu {\tilde \varphi}^{a*'}_+
                 + \frac{Z_6}{m} \partial_\mu h
                 {\tilde \varphi}^{a*'}_+
                 + 2 Z_{11} {\tilde \varphi}^{0*'}_+ A^a_\mu \, , \nonumber \\
&&
 {\tilde \varphi}^{a*}_+ = Z_4  {\tilde \varphi}^{a*'}_+ \, , ~~~~  {\tilde \varphi}^{0*}_+ = Z_8  {\tilde \varphi}^{0*'}_+ \, , \nonumber \\
&&
{\tilde b}^{a*} = - Z_7  {\tilde \varphi}^{a*'}_+ \tilde e^0
                  + Z_7 \varepsilon^{abc}   {\tilde \varphi}^{b*'}_+ \tilde e^c
                  - Z_{12}  {\tilde \varphi}^{0*'}_+ \tilde e^a + Z_m {\tilde b}^{a*'}  \, , \nonumber \\
&&
 {\tilde b}^{0*} = Z_7 {\tilde \varphi}^{a*'}_+ \tilde e^a -
                  Z_{12}  {\tilde \varphi}^{0*'}_+ \tilde e^0 + Z_m {\tilde b}^{0*'} \, , \nonumber \\
&&
h^* = -\frac{Z_6}{m} {\tilde \varphi}^{a*'}_+ \partial A^a
- \frac{Z_6}{m}  \partial_\mu {\tilde \varphi}^{a*'}_+ A^{a\mu}
+ \frac{Z_9}{m} \square {\tilde \varphi}^{0*'}_+ \nonumber \\
&&
~~~~~~ - 2\frac{Z_{10}}{m^2} \square h {\tilde \varphi}^{0*'}_+
-2\frac{Z_{10}}{m^2} \partial^\mu  {\tilde \varphi}^{0*'}_+ \partial_\mu h
+ Z_m Z_3 h^{*'} \, .
\label{app.1}
\end{eqnarray}

\section{Solution of the Linearized Functional Identity}\label{app.B}

This technical Appendix is devoted to the proof that all the divergences
can be recursively removed by a suitable choice of the parameters
$Z_g, Z_m, Z_\alpha$ and the field redefinition constants $Z_j, j=1,\dots,12$.
For that purpose
suppose that the subtraction of the divergences
has been performed up to order $n-1$
in the loop expansion while preserving
the  global
$SU(2)$ invariance and eq.(\ref{sc.1}).
Then at order $n$ eq.(\ref{sc.1}) gives
\begin{eqnarray}
&& {\cal S}_0(\G^{(n)}) \equiv
\int d^4x \, \sum_\Phi \Big ( \frac{\delta \G^{(0)}}{\delta \Phi^*(x)}
\frac{\delta}{\delta \Phi(x)} +
\frac{\delta \G^{(0)}}{\delta \Phi(x)}
\frac{\delta}{\delta  \Phi^*(x)}
\Big ) \G^{(n)} =  \nonumber \\
&& ~~~~~~~~~~~
- \sum_{j=1}^{n-1}
\int d^4x \, \sum_\Phi
\frac{\delta \G^{(n-j)}}{\delta \Phi^*(x)}
\frac{\delta \G^{(j)}}{\delta \Phi(x)}
\label{comm.3}
\end{eqnarray}
The second line of the above equation is finite since
it contains only lower order terms which have
already been subtracted. Hence
one gets the following equation for the divergent
part $\G^{(n)}_{div}$ at order $n$
\begin{eqnarray}
{\cal S}_0 (\G^{(n)}_{div}) = 0 \, .
\label{comm.4}
\end{eqnarray}
The action of ${\cal S}_0$ on the fields is the same
as that of $\delta$:
\begin{eqnarray}
{\cal S}_0 \Phi = \frac{\delta \G^{(0)}}{\delta \Phi^*} =
\delta\Phi \, .
\label{comm.5}
\end{eqnarray}
Moreover ${\cal S}_0$  acts on the antifield
$\Phi^*$ by mapping it into the classical e.o.m.
of $\Phi$, namely
\begin{eqnarray}
{\cal S}_0 \Phi^* = \frac{\delta \G^{(0)}}{\delta \Phi} \, .
\label{comm.6}
\end{eqnarray}
One can prove
in the usual fashion \cite{Gomis:1994he}
that ${\cal S}_0$ is nilpotent. This follows
from the nilpotency of $\delta$ and the validity of
the functional identity eq.(\ref{sc.1}) for $\G^{(0)}$.

\medskip
We are now going to prove that the divergences in $\G^{(n)}_{div}$
can be reabsorbed by a shift of the factors
$Z_g,Z_m,Z_\alpha$ and $Z_j$, $j=1,\dots,12$ which appear in $\hat
\G$. For that purpose we notice that the most general solution to
eq.(\ref{comm.4}) can be written as
\begin{eqnarray}
\G^{(n)}_{div} = A + {\cal S}_0 B
\label{comm.7}
\end{eqnarray}
where $A$ cannot be presented in the form ${\cal S}_0 C$, with $C$ a
local functional. Note that as $\Gamma^{(n)}$ is a Lorentz
invariant functional, the functionals $A$ and $B$ also possess this
invariance.

A convenient strategy for deriving the most general solution
eq.(\ref{comm.7}) can be described as follows.
We see that the Yang-Mills field-strength squared
\begin{eqnarray}
A = -\frac{a^{(n)}}{4} \int d^4x \,  G_{a\mu\nu} G_a^{\mu\nu} \, ,
\label{cofv.2}
\end{eqnarray}
where the divergent coefficient $a^{(n)}$ is unconstrained
by the symmetries, is a solution of the $A$-type.
The r.h.s. of
eq.(\ref{cofv.2}) is obviously also invariant under the global $SU(2)$
symmetry.

We must now address the question of whether other type-$A$ solutions exist.
One way to solve this problem is to compute the cohomology
$H_{{\cal F}}({\cal S}_0)$ \cite{HCohom}  of the nilpotent operator ${\cal S}_0$
in the space ${\cal F}$ of Lorentz- and global $SU(2)$-invariant local functionals
with dimension bounded by the power-counting.
$H_{\cal F}({\cal S}_0)$ is defined as the quotient of the latter functional space w.r.t. to the equivalence
relation
\begin{eqnarray}
X \sim Y \Leftrightarrow {X}-{Y}={\cal S}_0(C)
\label{equiv}
\end{eqnarray}
for some  Lorentz- and global $SU(2)$-invariant local functional $C$.

Clearly if we are able to   prove that $H_{{\cal F}}({\cal S}_0)$
 reduces to the equivalence class
of the Yang-Mills field strength squared (\ref{cofv.2}),
 we have also established
that the only type-$A$-solution to eq.(\ref{comm.7}) is given by
eq.(\ref{cofv.2}).

In order to evalute $H_{{\cal F}}({\cal S}_0)$ we first perform
the following change of
variables
\begin{eqnarray}
(\tilde{\varphi}^a_+)' = \frac{g}{2m} (
\tilde e^a \tilde b^0
- \tilde{e}^0
\tilde{b}^a -\varepsilon^{abc}\tilde{e}^b \tilde{b}^c
) + \tilde{\varphi}^a_+\nonumber\\
(\tilde{\varphi}_0^+)' = \frac{g}{2m}( \tilde{e}^a \tilde{b}^a+\tilde{e}^0 \tilde{b}^0) +
\tilde{\varphi}^0_+ \, ,
\label{cofv}
\end{eqnarray}
which leads to the following ${\cal S}_0$-transforms
\begin{eqnarray}
&& {\cal S}_0  h = - \tilde{b}^0, ~~~ {\cal S}_0 ( \tilde{b}^0 )= 0 \nonumber \\
&& {\cal S}_0 \tilde e_a = (\tilde{\varphi}^a_+)'  \, , ~~~
{\cal S}_0 (\tilde{\varphi}^a_+)'  = 0 \nonumber \\
&& {\cal S}_0 \tilde e_0 = (\tilde{\varphi}^0_+)'  \, , ~~~
{\cal S}_0 (\tilde{\varphi}^0_+)'  = 0 \, .
\label{cofv.1}
\end{eqnarray}
The reason for carrying out such a field redefinition stems from the properties
of the so-called doublet variables.
A pair of variables $u,v$ such that
${\cal S}_0 u = v \, , {\cal S}_0 v =0$
is called a  ${\cal S}_0$-doublet.
There is a very general theorem about
doublets for nilpotent linear operators
\cite{Barnich:2000zw, Quadri:2002nh} which states that
 the dependence on $u,v$
can only happen via the term ${\cal S}_0B$.

I.e. the cohomology of ${\cal S}_0$ is isomorphic
to that of the restriction of ${\cal S}_0$
to the subspace without doublets
(and their antifields).
As a check of this fact it can be verified by explicit computation
that all the terms in $A'_{ef}$
depending on the  variables
$h, -\tilde{b}^0$, $ \tilde e_a, (\tilde{\varphi}^a_+)'$,
$\tilde e_0, (\tilde{\varphi}^0_+)'$,
pairing into doublets according to eq.(\ref{cofv.1}),
can indeed be written as the $\delta$-variation of a  local
Lorentz-invariant functional.

In the subspace where the doublets
(and their antifields) are dropped,
the action of ${\cal S}_0$ on $A_{a\mu}$, $\tilde
b_a$ is the same as the standard gauge BRST transformation for the
gauge group $SU(2)$, once one identifies the $SU(2)$ BRST ghosts
with $\frac{1}{m} \tilde b_a$. As a consequence, the dependence on
the antifields $A^{*a}_\mu, \tilde b^{a*}$ is also confined in the
${\cal S}_0B$-functional \cite{HCohom}.
So we conclude that
the cohomology
of ${\cal S}_0$ is isomorphic to the one of the SU(2) Yang-Mills
theory, as one should expect.
This result holds irrespectively of the dimensions of the local operators involved.

In the space of operators of dimension $\leq 4$ the only
element of this cohomology is the Yang-Mills field strength squared
in eq.(\ref{cofv.2}). Therefore we see that there are no further
type-A solution to be considered and we can now proceed to enumerate all
the invariants of the second type ${\cal S}_0B$.

They fall into two classes: those which do not involve
the antifields and those which depend on the antifields.
By imposing the dimensional bounds dictated by power-counting and the residual global SU(2) invariance
three invariants of the first class arise.
The first one is
\begin{eqnarray}
{\cal J}_t = \int d^4x \, {\cal S}_0 \tilde e_0 =
\int d^4x \, \Big (
 \frac{g}{2m} ( \tilde{e}^a \tilde{b}^a+ \tilde{e}^0 \tilde{b}^0)+
\tilde{\varphi}^0_+
\Big ) \, .
\label{tad.1}
\end{eqnarray}
It controls the tadpole for $\tilde{\varphi}^0_+$ and never arises
in dimensional regularization.
The other two can
be conveniently described in terms of
the following operator insertions
\begin{eqnarray}
&& {\cal J}_m =
\delta Z_m^{(n)}
 \frac{\partial}{\partial Z_m} \hat \G \, , ~~~~
{\cal J}_\alpha =
\delta Z_\alpha^{(n)}
 \frac{\partial}{\partial Z_\alpha} \hat \G \, ,
\label{ins.1}
\end{eqnarray}
where $\delta Z_m^{(n)}, \delta Z_\alpha^{(n)}$
are divergent coefficients of order $n$.

\medskip
The invariants of the ${\cal S}_0 B$-type involving the antifields
are of the form (no sum over $\Phi$)
\begin{eqnarray}
z^{(n)}_j \int d^4x \, {\cal S}_0 ( \Phi^* F_j(\Phi)) =
z^{(n)}_j \int d^4x \Big ( F_j(\Phi) \frac{\delta \G^{(0)}}{\delta \Phi}
- \Phi^* \frac{\delta F_j}{\delta \Phi} {\cal S}_0 \Phi \Big )
\label{comm.8}
\end{eqnarray}
where again the divergent coefficient $z^{(n)}$ is
of order $\hbar^n$.

The possible $F_j$'s, $j=1,\dots,12$
in eq.(\ref{comm.8}) are constrained
by the rigid symmetries of the theory,
quantum numbers and power-counting
and have the same structure as the
corresponding terms in eq.(\ref{sc.3}).

$Z_m$,$Z_\alpha$,$Z_g$ and $Z_j$, $j=1,\dots,12$
are formal power series in $\hbar$ of the general form
$Z= 1 + \sum_{j=1}^\infty Z^{(j)}$.
Their coefficients have been fixed up to order
$n-1$ due to the recursion assumption.

We must now prove that their $n$-th order
coefficients can be chosen in such a way to remove
the divergences in eq.(\ref{comm.7}).

For that purpose it is convenient to
redefine $A^a_\mu \rightarrow Z_g g A^a_\mu$. Then the coefficient
of the Yang-Mills field-strength squared in $\hat \G$ becomes
$-\frac{1}{4 Z_g^2 g^2}$ and the term $A$ in eq.(\ref{cofv.2}) is
reabsorbed by choosing  $Z_g^{(n)} = \frac{a^{(n)}}{2}$. The
terms in eq.(\ref{ins.1}) are recovered by
choosing  $Z_m^{(n)} = - \delta Z^{(n)}_m$ and
$Z_\alpha^{(n)} = - \delta Z^{(n)}_\alpha$.

 We now move to the terms in eq.(\ref{comm.8}).
They can be reabsorbed by setting $Z_j^{(n)} = - z_j^{(n)}$,
since by using eqs.(\ref{sc.3})
and (\ref{app.1})  in eq.(\ref{sc.5}) one gets at the first
non-vanishing order in $Z_j^{(n)}$
\begin{eqnarray}
&& Z_j^{(n)} \int d^4x \, \Big (
F_j(\Phi) \frac{\delta \G^{(0)}}{\delta \Phi}
\Phi - \Phi^* \frac{\delta F_j}{\delta \Phi}
{\cal S}_0 \Phi \Big ) + O(\hbar^{n+1}) = \nonumber \\
&& ~~~~~~
- z_j^{(n)} \int d^4x \, \Big (
F_j(\Phi) \frac{\delta \G^{(0)}}{\delta \Phi}
\Phi - \Phi^* \frac{\delta F_j}{\delta \Phi}
{\cal S}_0 \Phi \Big ) + O(\hbar^{n+1})
\label{comm.new.1}
\end{eqnarray}

Once the $n$-th order divergences have been removed,
the procedure can be recursively applied.
In fact  $\hat \G$
obeys eq.(\ref{sc.2}) and thus the functional identity will
be fulfilled at the order $n+1$.
Since the divergences have been subtracted
up to order $n$, eq.(\ref{comm.3}) holds
at order $n+1$ and the argument can be repeated.

So indeed the ultraviolet divergencies generated by the
interaction may be removed by changing the parameters entering the
classical action, expressed in terms of the redefined fields. Hence
the renormalized theory is also unitary in the subspace including
only three dimensionally transversal components of the Yang-Mills
field.


\begin{thebibliography}{99}
%
\bibitem{FP}
  L.~D.~Faddeev and V.~N.~Popov,
  Phys.\ Lett.\  B {\bf 25} (1967) 29;
  \quad Perturbation theory for gauge
  invariant fields. Preprint ITP, Kiev 1967.

\bibitem{DW}
  B.~S.~DeWitt,
  Phys.\ Rev.\  {\bf 160} (1967) 1113, 1195.

\bibitem{VG}
  V.~N.~Gribov,
   Nucl.\ Phys.\  B {\bf 139} (1978) 1.

\bibitem{IS}
  I.~M.~Singer,
  Commun.\ Math.\ Phys.\  {\bf 60} (1978) 7.

\bibitem{DZ}
  D.~Zwanziger,
    Nucl.\ Phys.\  B {\bf 321} (1989) 591;
    Nucl.\ Phys.\  B {\bf 323} (1989) 513.

\bibitem{KO}
  T.~Kugo and I.~Ojima,
  Prog.\ Theor.\ Phys.\ Suppl.\  {\bf 66} (1979) 1.

\bibitem{Sl1}
  A.~A.~Slavnov,
  JHEP {\bf 0808} (2008) 047
  [arXiv:0807.1795 [hep-th]].

\bibitem{Sl2}
  A.~A.~Slavnov, Theor.\ Math.\ Phys.\ {\bf 161} (2009) 204.

\bibitem{Bar}
 R.~N.~Baranov, Theor.\ Math.\ Phys. \ {\bf 161} (2009) 37.

\bibitem{VT}
  B.~L.~Voronov I~V~Tyuin, Theor.\ Math.\ Phys.\ {\bf 50} (1982) 218; {\bf 52} (1982) 628.

\bibitem{ZJ}
 J.~Zinn-Justin,
 Renormalization of Gauge Theories.
Lectures given at Int. Summer Inst. for Theoretical Physics, Jul 29 - Aug 9, 1974, Bonn, West Germany.
Published in Bonn Conf.1974:2 (QCD161:I83:1974)

\bibitem{Troost:1989cu}
  W.~Troost, P.~van Nieuwenhuizen and A.~Van Proeyen,
  Nucl.\ Phys.\  B {\bf 333} (1990) 727.


\bibitem{Betal}
  Ph.~Boucaud, J.~P.~Leroy, A.~Le Yaouanc, J.~Micheli, O.~Pene and J.~Rodriguez-Quintero,
  JHEP {\bf 0806} (2008) 012
  [arXiv:0801.2721 [hep-ph]];
  JHEP {\bf 0806} (2008) 099
  [arXiv:0803.2161 [hep-ph]].
%

\bibitem{ABP}
  A.~C.~Aguilar, D.~Binosi and J.~Papavassiliou,
  Phys.\ Rev.\  D {\bf 78} (2008) 025010
  [arXiv:0802.1870 [hep-ph]].

\bibitem{FMP}
  C.~S.~Fischer, A.~Maas and J.~M.~Pawlowski,
  Annals Phys.\  {\bf 324} (2009) 2408
  [arXiv:0810.1987 [hep-ph]].

\bibitem{CM}
  A.~Cucchieri and T.~Mendes,
  Phys.\ Rev.\ Lett.\  {\bf 100} (2008) 241601
  [arXiv:0712.3517 [hep-lat]].

\bibitem{BIMS}
  I.~L.~Bogolubsky, E.~M.~Ilgenfritz, M.~Muller-Preussker and A.~Sternbeck,
  Phys.\ Lett.\  B {\bf 676} (2009) 69
  [arXiv:0901.0736 [hep-lat]].

\bibitem{Gomis:1994he}
  J.~Gomis, J.~Paris and S.~Samuel,
  Phys.\ Rept.\  {\bf 259} (1995) 1
  [arXiv:hep-th/9412228].

\bibitem{Barnich:2000zw}
  G.~Barnich, F.~Brandt and M.~Henneaux,
  Phys.\ Rept.\  {\bf 338} (2000) 439
  [arXiv:hep-th/0002245].

\bibitem{HCohom}
  G.~Barnich and M.~Henneaux,
  Phys.\ Rev.\ Lett.\  {\bf 72} (1994) 1588
  [arXiv:hep-th/9312206];
  G.~Barnich, F.~Brandt and M.~Henneaux,
  Commun.\ Math.\ Phys.\  {\bf 174} (1995) 57
  [arXiv:hep-th/9405109];  Commun.\ Math.\ Phys.\  {\bf 174} (1995) 93
  [arXiv:hep-th/9405194].


\bibitem{Quadri:2002nh}
  A.~Quadri,
  JHEP {\bf 0205}, 051 (2002)
  [arXiv:hep-th/0201122].



\end{thebibliography}
\end{document}